\documentclass{aa}

\usepackage{graphicx}
\usepackage{times}

\newcommand{\pun}[1]{\mbox{\rm\,#1}} % Command used to write physical units
\newcommand{\abuhe}{\mbox{Y}}
\newcommand{\logg}{\ensuremath{\log g}}
\newcommand{\grav}{\ensuremath{g}}
\newcommand{\mlp}{\ensuremath{\alpha_{\mathrm{MLT}}}}
\newcommand{\mlpcm}{\ensuremath{\alpha_{\mathrm{CMT}}}}
\newcommand{\moh}{\ensuremath{[\mathrm{M/H}]}}
\newcommand{\senv}{\ensuremath{\mathrm{s}_{\mathrm{env}}}}

\newcommand{\spun}{\ensuremath{\mathrm{s}_0}}
\newcommand{\sstar}{\ensuremath{\mathrm{s}^\ast}}
\newcommand{\Teff}{\ensuremath{T_{\mathrm{eff}}}}
\newcommand{\tauross}{\ensuremath{\tau_{\mathrm{ross}}}}
\newcommand{\ttaurelation}{\mbox{T$(\tau$)-relation}}

\newcommand{\draftflag}{false}

\newcommand{\beq}{\begin{equation}}
\newcommand{\eeq}{\end{equation}}
\newcommand{\pdx}[2]{\frac{\partial #1}{\partial #2}}

\newcommand{\eref}[1]{\mbox{(\ref{#1})}}

\newcommand{\Vact}{\ensuremath{\nabla}}
\newcommand{\Vad}{\ensuremath{\nabla_{\mathrm{ad}}}}
\newcommand{\Veddy}{\ensuremath{\nabla_{\mathrm{e}}}}
\newcommand{\Vrad}{\ensuremath{\nabla_{\mathrm{rad}}}}

\newcommand{\cp}{\ensuremath{c_{\mathrm{p}}}}
\newcommand{\taueddy}{\ensuremath{\tau_{\mathrm{e}}}}
\newcommand{\vconv}{\ensuremath{v_{\mathrm{c}}}}
\newcommand{\Fconv}{\ensuremath{F_{\mathrm{c}}}}
\newcommand{\lmix}{\ensuremath{\Lambda}}
\newcommand{\Hp}{\ensuremath{H_{\mathrm{P}}}}
\newcommand{\Hptop}{\ensuremath{H_{\mathrm{P,top}}}}

\newcommand{\Teffref}{\ensuremath{\tilde{T}}}
\newcommand{\loggref}{\ensuremath{\tilde{g}}}

\begin{document}
\thesaurus{    06     % A&A Section 6: Form. struct. and evolut. of stars
           (02.03.3;  % convection
            02.08.1;  % hydrodynamics
            08.12.1;  % Stars: late-type
            08.05.3)} % Stars: evolution.
\title{A calibration of the mixing-length for solar-type stars based
         on hydrodynamical simulations}
\subtitle{I. Methodical aspects and results for solar metallicity}

\titlerunning{A calibration of the mixing-length for solar-type stars I.}
\authorrunning{Ludwig et al.}

\author{ Hans-G\"unter Ludwig\inst{1} \and Bernd Freytag\inst{1,2}
  \and Matthias Steffen\inst{3,2} } \offprints{Hans-G\"unter Ludwig}
  \institute{ Astronomical Observatory, %Niels Bohr Institute,
  Juliane Maries Vej 30, DK-2100 Copenhagen \O, Denmark 
  [hgl@astro.ku.dk, bf@astro.ku.dk] \and
  Institut f\"ur Theoretische Physik und Astrophysik der Universit\"at Kiel,
  D-24098 Kiel, Germany
%  WWW: http://www.astrophysik.uni-kiel.de/
\and
  Astrophysikalisches Institut Potsdam,
  D-14482 Potsdam, Germany,
  [MSteffen@aip.de]
}

\date{Received 28 September 1998; accepted date}

\maketitle

\begin{abstract}
Based on detailed 2D numerical radiation hydrodynamics (RHD)
calculations of time-dependent compressible convection, we have
studied the dynamics and thermal structure of the convective surface
layers of solar-type stars.  The RHD models provide information about
the convective efficiency in the superadiabatic region at the top of
convective envelopes and predict the asymptotic value of the entropy
of the deep, adiabatically stratified layers (Fig.~\ref{f:sstarhd}).
This information is translated into an effective mixing-length
parameter~\mlp\ suitable to construct standard stellar structure
models.  We validate the approach by a detailed comparison to
helioseismic data.

The grid of RHD models for solar metallicity comprises 58 simulation
runs with a helium abundance of $\abuhe=0.28$ in the range of
effective temperatures $4300\pun{K}\leq\Teff\leq 7100\pun{K}$ and
gravities $2.54\leq\logg\leq 4.74$.  We find a moderate, nevertheless
significant variation of \mlp\ between about 1.3 for F-dwarfs and 1.75
for K-subgiants with a dominant dependence on \Teff\
(Fig.~\ref{f:mlp}).  In the close neighbourhood of the Sun we find a
plateau where \mlp\ remains almost constant.  The internal accuracy of
the calibration of \mlp\ is estimated to be $\pm 0.05$ with a possible
systematic bias towards lower values. An analogous calibration of the
convection theory of Canuto \&\ Mazzitelli (1991, 1992; CMT) gives a
different temperature dependence but a similar variation of the free
parameter (Fig.~\ref{f:mlpcm}).

For the first time, values for the gravity-darkening exponent $\beta$
are derived independently of mixing-length theory: $\beta = 0.07\ldots
0.10$.

We show that our findings are consistent with
constraints from stellar stability considerations and provide compact
fitting formulae for the calibrations.
\end{abstract}

\section{Introduction}

The structure and evolution of late-type stars is intimately related
to convective transport processes taking place in their outer
envelopes.  Although the underlying physical principles are well
known, the non-linear and non-local character of the equations
describing the convective motions in a radiating, partially ionized
fluid has hampered the development of a closed analytical theory.
Hitherto standard stellar structure models include the convective
energy transport in the framework of mixing-length theory (MLT).
Despite its heuristic nature MLT has proven to be rather successful
and is still the working horse in stellar structure modeling.  During
recent years much effort has been put into the refinement of the
theoretical description of microphysical stellar plasma properties ---
the opacity and the equation of state.  To fully benefit
from these improvements, an accompanying development of the description
of hydrodynamical transport properties appears to be necessary, in the
first place aiming at a better understanding of the convective energy 
transport.  In this paper we
report on such an effort relying on direct numerical simulations of
convective flows in solar-type stars.

Primarily due to the heavy demands on computational resources, RHD
simulations have concentrated very much on the Sun, and only rather few
models have been constructed for other objects (Nordlund 1982, Steffen
et al. 1989, Nordlund \&\ Dravins 1990, Ludwig et al. 1994, Freytag et
al. 1996).  This sparseness has prevented a broader application of
results of RHD simulations to the modeling of stellar atmospheres
and evolution.  We try to overcome this limitation in this work by
computing a {\em grid of RHD models\/}, allowing a systematic study of
the characteristics of convective flows in a larger region of the
Hertzsprung-Russell diagram (HRD).  We point out that the RHD simulations 
have now reached a
level of sophistication far beyond idealized, simple numerical
experiments. We consider it an important aspect of our work that it
presents a {\em quantitative study\/} and --- as such --- puts emphasis on
the estimation of the achieved accuracy.  Aiming at an application in the 
field of stellar evolution, we concentrate on the evaluation of one 
structural quantity, namely the entropy of the adiabatically stratified 
layers deep in the convective stellar envelope. In the following we shall
refer to this entropy as \senv.  Its relevance for stellar structure
stems from the fact that it strongly influences the radius of a star.
To obtain handy numbers, all entropy values in this paper --- unless
stated otherwise --- are given in units of $\spun=10^9\pun{erg/g/K}$.
To link our results more directly to standard stellar structure
modeling we translate \senv\ into an equivalent mixing-length
parameter.

Some words concerning our nomenclature: we use the term ``solar-type''
for stars with extended convective envelopes where the thickness of the
superadiabatic layers at the top of this envelope is small in
comparison to the stellar radius.
``Grey'' radiative transfer means that frequency-independent (mean)
opacities were used in the computation of the radiation field.  The
opacities still depend on temperature and density and include
contributions from spectral lines.

In the paper we proceed as follows: we start with methodical aspects
describing our hydrodynamical models, the basic idea and the procedure
to derive \senv, and the translation of \senv\ into an equivalent
mixing-length. We continue with the validation of our method by
showing that we are able to predict the solar structure derived from
helioseismic measurements within small uncertainties. We then present
the calibrations of MLT and CMT. We discuss 
the application of our results to stellar modeling, point out consequences 
of stellar stability considerations, present a derivation of the 
gravity-darkening exponent, and contrast our approach with others. 
We conclude with future perspectives. In the appendix we provide some 
auxiliary data helping to utilize our findings in stellar structure models.

\section{Methodical aspects}

\subsection{Hydrodynamical models of solar-type surface convection} 

We have obtained detailed 2-dimensional models of the surface layers
of solar-type stars from extensive numerical simulations solving the
time-dependent, non-linear equations of hydrodynamics for a stratified
compressible fluid. The calculations take into account a realistic
equation-of-state (EOS, including the ionization of H and He as well as
formation of $\mathrm{H}_2$-molecules) and use an elaborate scheme to
describe multi-dimensional, non-local, frequency-dependent radiative
transfer. Similar to classical model atmospheres, the hydrodynamical
models are characterized by effective temperature~\Teff, acceleration
of gravity~\logg, and chemical composition.  They include the
photosphere as well as part of the subphotospheric layers, with an
{\em open lower boundary\/}, allowing a free flow of gas out of and
into the model.  A fixed specific entropy~\sstar\ is (asymptotically)
assigned to the gas entering the simulation volume from below. The value
adopted for \sstar\
uniquely determines the effective temperature of the hydrodynamical
model. 
For details about the physical assumptions, numerical method
and characteristics of the resulting convective flows
see Ludwig et al. (1994)
and Freytag et al. (1996).

\subsection{From the surface to the base of a convection zone}

\begin{figure*}[!t]
\resizebox{\hsize}{!}{\includegraphics[draft = \draftflag]%
{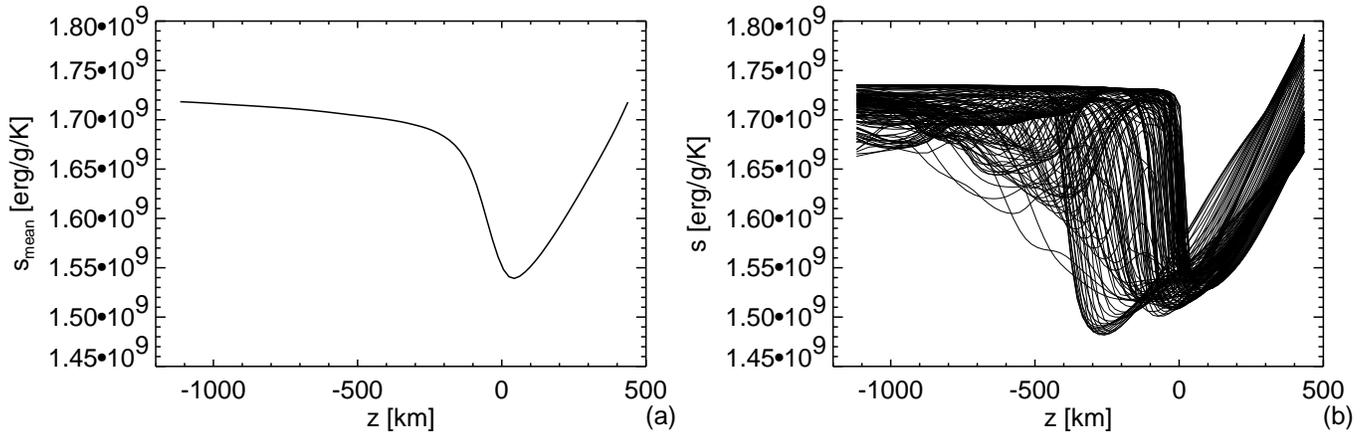}}
\caption[]{%
Depth dependence of the entropy in the solar surface layers as obtained from
hydrodynamical simulations ($\Teff=5770\pun{K}$, $\logg=4.44$, model
code L71D07) performed on a 140 (x) by 71 (z)
grid with frequency-dependent radiative transfer. The mean entropy
(horizontal and temporal average) is shown in panel~(a), spatially resolved
entropy profiles in panel~(b). Geometrical height zero corresponds to
$\tau_{\mathrm{Ross}}=1$. 
Note that the model comprises only the uppermost part of the
200\pun{Mm} deep solar convective zone.
\label{f:splot}
} % end of caption
\end{figure*}

Figure~\ref{f:splot}a shows the {\em mean\/} entropy as a function
of depth obtained from a hydrodynamical granulation model of the Sun by
averaging over horizontal planes and over time. As in this example,
our models in general do not extend deep enough to include those
layers where the mean stratification of the convection zone becomes
adiabatic. While the mean entropy stratification of the hydrodynamical
models does not permit a direct determination of the entropy
corresponding to the adiabat of the deep convection zone, the {\em
spatially resolved\/} entropy profiles contain additional information.
Figure~\ref{f:splot}b displays the entropy profiles for an arbitrary
instant of the sequence from which the mean stratification in
Fig.~\ref{f:splot}a was computed.  The granular convection pattern
at the surface of solar-type stars is formed by broad hot upflows
accompanied by concentrated cool downdrafts.  Figure~\ref{f:splot}b
shows a remarkable entropy plateau in the subsurface layers,
indicating that --- in contrast to the narrow downdrafts --- the gas
in the central regions of the broad ascending flows is still thermally
isolated from its surroundings. Neither radiative losses nor
entrainment by material of low entropy can produce significant
deviations from adiabatic expansion until immediately below the
radiating surface layers. The height of the entropy plateau is
essentially independent of time and corresponds to \sstar.

% --- Spatially resolved/unresolved entropy plot

We suggest that \sstar\ may be identified directly with the
entropy~\senv\ of the deep, adiabatic convective layers:

\beq
\senv = \sstar.
\eeq

\noindent 
This idea has
been put forward by Steffen (1993) and Ludwig et al. (1997) and is
based on the qualitative picture of solar-type convection zones
proposed by Stein \&\ Nordlund (1989) (hereafter referred to as ``SN
scenario'') which is fundamentally different from MLT
assumptions. According to this scenario the downdrafts continue all
the way from the surface to the bottom of the convection zone, merging
into fewer and stronger currents at successively deeper levels. The
flow closes only near the base of the convective envelope. Most of the
gas elements starting from the bottom of a deep convection zone
overturn into neighbouring downflows before reaching the surface. Only
a very small fraction of gas continues to the surface, reaching the
layers corresponding to the location of the lower boundary of our
hydrodynamical models essentially without entropy losses, following an
adiabat almost up to the visible surface. Hence, \sstar\ obtained from
the simulations is the entropy of the warm, ascending gas throughout
the convection zone.  This, in turn, is very nearly equal to the mean
(horizontally averaged) entropy~\senv\ near the base of the convection
zone because (i) the downflows are markedly entropy-deficient only
near the surface and become continuously diluted by overturning
entropy-neutral gas as they reach greater depths, and (ii) the
fractional area occupied by the downdrafts decreases with depth.

We note that for this investigation it is actually irrelevant whether
the downflows possess a plume-like character or take place in
another form. The essential point is that sinking material does not
locally affect the entropy of rising material by draining heat from
it. The role of the downflows is reduced to a dynamical one: while 
sinking downwards they simply displace buoyancy-neutral material
and push it upwards.

\subsection{Functional dependence of \senv\ on stellar parameters}

When considering the MLT picture and the SN scenario it is interesting
to ask what stellar properties determine \senv\ or --- in other
words --- what coordinates are appropriate to describe its functional
dependence in the HRD.  We argued that hydrodynamical models of the
surface layers are able to provide information about \senv. The models
are characterized by the {\em atmospheric parameters\/} and
consequently we describe \senv\ in terms of them. Whether the standard
atmospheric parameters \Teff\ and \logg\ are the most suitable
coordinates is not clear.  One might speculate that e.g. the surface
opacity is a physically more relevant quantity.  Nevertheless, for
solar-type stars the conditions at the stellar surface govern the
global envelope structure and the standard atmospheric parameters are
suitable coordinates to parameterize them.  They have the advantage
that they are external control parameters and not part
of the solution of the problem.  For solar-type stars with
superadiabatic regions which are thin in comparison to the stellar
radius we expect the same qualitative behaviour irrespective of whether
we consider the MLT picture or the SN scenario.  Global stellar
parameters (mass, radius, or age) play only an indirect role for the
entropy jump.  The situation changes when the size of the granular
cells or the thickness of the superadiabatic layer become comparable
to the stellar radius.  This might happen in giants and one has to
account for effects introduced by the global stellar structure,
i.e. sphericity effects.

\subsection{\mlp\ from envelope models}

\begin{figure}
\resizebox{\hsize}{!}{\includegraphics[draft=\draftflag]
{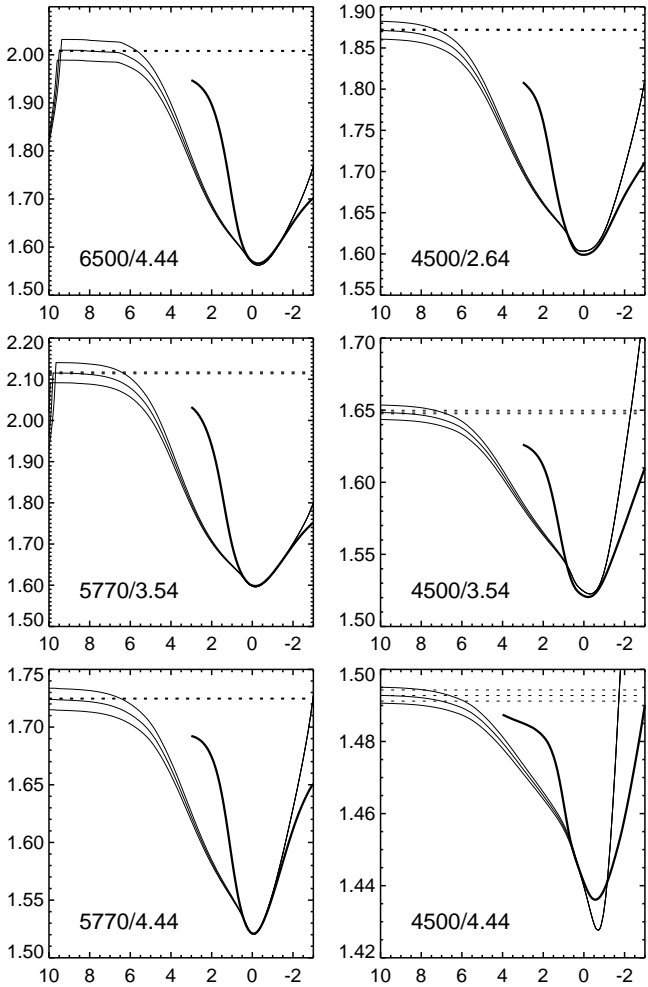}}
\caption[]{%
Representative examples of entropy profiles (in units of
$\spun=10^9\pun{erg/g/K}$) as a
function of Rosseland optical depth $\log\tauross$. Each panel shows
the $\tau$-averaged entropy of a RHD model (thick solid line) with 
envelope models (thin solid lines) for the matching \mlp\ and $\mlp\pm
0.05$ for a given $\Teff/\logg$ combination. 
In all models the radiative transfer was treated in the grey
approximation.
The dotted horizontal
lines indicate \sstar\ of the RHD model and its temporal
RMS-fluctuations. Note the significantly different entropy jumps.
\label{f:splots}
} % end of caption
\end{figure}

Although \senv\ allows the construction of the envelope structure with the
necessary precision, it is of interest to translate this quantity to an
equivalent \mlp\ since this parameter is conventionally used in stellar 
structure models.  For this purpose we computed grids of standard stellar
envelope models (not subject to central boundary conditions) based on
MLT, covering the relevant range of effective temperature and
gravity and assuming $\abuhe=0.28$ and solar metallicity for the
chemical composition. In these models \mlp\ was a free
parameter. For given stellar parameters \mbox{(\Teff, \logg)} we obtained
\senv\ as a function of \mlp. By matching \senv\ from the RHD models
we deduced the corresponding \mlp. 
Figure~\ref{f:splots} illustrates this procedure for a number of
representative models. 
Each panel shows the entropy profile of the envelope model matching
\senv\ from a RHD simulation.
For comparison two further envelope models are plotted with \mlp\
varied by $\pm 0.05$ with respect to the matching one, as well as an
estimate of the uncertainty of \senv\ stemming from the temporal
fluctuations of \sstar\ which we observe during the simulation run.
Although at first glance the matching is a
straightforward procedure, it is important to take care of a number of 
fine points in order to ensure a unique and well-defined calibration of 
\mlp. We discuss these points in the following.

Concerning the {\em opacities\/} and {\em EOS\/} we tried to stay as
close as possible to the physical description used in the RHD models.
Due to this differential procedure, systematic uncertainties are substantially 
reduced.  In the RHD models ATLAS6 opacities according Kurucz (1979)
are adopted.  These opacities contain no contribution from molecular
lines.  Since the envelope models reach much deeper than the RHD
models we supplemented the ATLAS6 opacities by OPAL opacities for
higher temperatures (Rogers \& Iglesias 1992). Care was taken that the
chemical mixture assumed in both data sets was as similar as possible.
In the envelope models we used the RHD~EOS throughout.  It is a simple
EOS taking into account the ionization equilibria of hydrogen and
helium according to the Saha-Boltzmann equations.

The RHD~EOS has the advantage of great smoothness and computational
efficiency.  It might appear that this description is too crude,
especially for the envelopes of cooler objects where non-ideal effects
become more pronounced. However, this simplification is tolerable since the
entropy profile, which is the major concern here, is not sensitively
dependent on the EOS.  The entropy gradient is proportional to the
difference between actual and adiabatic temperature gradient which in MLT is
related to the convective flux.  Hence, the constraint that a nominal total
flux has to be transported essentially fixes the entropy gradient 
irrespective of the EOS employed, while the
temperature profile is noticeably affected by the choice of the EOS. 

Following the conventional way to treat the atmosphere in stellar
structure models we prescribe a certain {\em \ttaurelation\/}
representing the temperature run in the optically thin regions.  Its
significance for the envelope structure emerges from the fact that it
determines the level of the entropy minimum in the deeper atmosphere
where the stratification becomes convectively unstable (the starting level 
for the entropy jump).  In evolutionary
calculations various descriptions of the atmosphere are commonly used,
differing in the level of sophistication between simple analytical or
empirical {\ttaurelation}s and full-fledged model atmospheres.  Here
we chose a \ttaurelation\ that mimics closely the average atmospheric
structure of the hydrodynamical models.  We reproduce in particular
the atmospheric entropy minimum found in the hydrodynamical models.
The hydrodynamical models predict an average temperature structure in
the deeper photosphere (around $-2<\log\tauross<0$) that closely
resembles a stratification in radiative equilibrium.  We
emphasize that this statement refers to an averaging of the
temperature on surfaces of constant optical depth in the RHD models.
This averaging procedure preserves the radiative flux properties of the 
RHD models to a good approximation (see Steffen et al. 1995), and hence
appears suitable to characterize their atmospheric structure in the
present context since radiative surface cooling plays the dominant
role for the convection driving.

In practice, we used an analytical fit to the exact grey
\ttaurelation\ for an atmosphere in radiative equilibrium according
Uns\"old (1955) in envelope models intended for comparison with RHD
models adopting grey radiative transfer.  From the
small set of RHD models adopting frequency-dependent radiative
transfer we derived the average \ttaurelation\ and constructed an
analytical fit to it. Later in this paper we shall discuss the role of
grey versus frequency-dependent radiative transport
in the RHD models, demonstrating the need for a clean distinction between 
both types of {\ttaurelation}s.

The influence of the \ttaurelation\ grows as the entropy jump itself
decreases, so it is largest for models at low effective
temperature and high gravitational acceleration.
In earlier papers on the subject (Ludwig et al. 1996, 1997) we used
{\ttaurelation}s which were not selected on a strictly differential
basis, leading to systematic differences between the older and the
present calibration of \mlp.
Figure~\ref{f:splots} illustrates the match of the atmospheric entropy
minima.  (The range in optical depth over which the RHD models are
plotted do not represent the full extension of the RHD models.)  Our
present procedure gives now an improved match with the largest
deviation found for the \mbox{($4500, 4.44$)} model. We expect that a
further improved fit of the atmospheric stratification of the RHD
model would lead to an increase of \mlp.  But note that using the
\mlp\ derived from models in this region of the HRD would produce a
consistent \senv\ in a stellar structure model as long as it is basing
on a description of the stellar atmosphere by a \ttaurelation\
resembling an atmosphere in radiative equilibrium.

As a side-issue the Fig.~\ref{f:splots} 
demonstrates that in RHD models the higher
atmospheric layers are systematically cooler than in an atmosphere in
radiative equilibrium.  This is due to the situation that in a
convective atmosphere the temperature is governed by the competition
between adiabatic cooling of overshooting material and heating by the
radiation field.  Quantatively, our grey RHD models overestimate the
temperature reduction since they do not include the important heating
by spectral lines in the higher layers. However, as we will see in the
next section this is of minor importance for the determination of
\mlp. 
The deviations between average RHD stratification and envelope models
underline that our calibrated \mlp\ only fits the asymptotic entropy
and the entropy jump. The overall stratification is not well matched
by an envelope model with the fitted \mlp.

Effects of {\em turbulent pressure\/} are ignored in the envelope
models.  This is not only done for computational convenience but also
reflects our opinion that the recipes provided for its treatment in
the framework of MLT overestimate the effects, doing more harm
then good when included.  In MLT the convective motions are restricted
to the unstable layers according to the Schwarzschild criterion.  At
the upper boundary of a surface convective zone this produces a sharp
decline of the convective velocity amplitude and a correspondingly
large gradient of the turbulent pressure that alters the hydrostatic
structure of a model significantly. In contrast, hydrodynamical models
(see Freytag et al. 1996) predict a rather smooth run of the velocity
amplitude in that region with a correspondingly smaller gradient of
the turbulent pressure.  Clearly, the inclusion of turbulent
pressure in 1D envelope models requires a proper treatment of overshooting. 

Finally, it is important to realize that a unique formulation of
MLT does not exist. Rather, {\em different
versions of MLT\/} are in use. In the following we refer to the
version of MLT originally given by B\"ohm-Vitense (1958). To further
elaborate this point, we explicitly present details of our MLT
implementation together with commonly used other
formulations in Appendix~A.

% --- fit examples

\section{The solar benchmark}
\label{s:sbm}

The natural benchmark for the scenario described above is of course the
Sun. Not only are its global stellar parameters known with
exceptional accuracy, but helioseismology has provided us with detailed
information about the structure of the solar interior.  Recent
astrophysical convection theories were developed in the context of
stellar evolution and tests of these theories were devised within that
framework, in particular by studying their effects on the evolution of the
Sun (Lydon et al. 1992, 1993a; Canuto \&\ Mazzitelli 1991, 1992; Canuto et
al. 1996). In our opinion, a more direct way for a validation of such
theories is to look at the present solar structure without taking
recourse to evolutionary calculations.  This avoids the possibility
that evolutionary changes of the solar structure interfere with model
properties that are used for the assessment of the accuracy of the
convection theory under consideration.

In the following we shall compare our RHD model predictions for \senv\
with related helioseismic measurements.  On an absolute scale any
deviation from the actual physical situation, e.g. in composition,
shows up as mismatch between observations and theoretical predictions
without being necessarily related to flaws in the convection model.
In our solar models we paid particular attention to the role of the
{\em helium content~\abuhe\/} and {\em details of the radiative
transfer\/}.

\begin{table}
\begin{flushleft}
\caption[]{%
Results from hydrodynamical (RHD) models and helioseismology for
different helium abundances.  \abuhe: helium abundance, \senv: entropy in
the adiabatically stratified layers in units of $\spun=10^9\pun{erg/g/K}$,
 $N_\nu$: number of frequency
bands for modeling the radiative transfer, \mlp: mixing-length
parameter. The data for $\abuhe=0.246$ are obtained by linear interpolation
between the data for $\abuhe=0.24$ and $0.28$.}
\begin{tabular}{lllll}
\hline\noalign{\smallskip}
Source           & \abuhe& $N_\nu$ & $\senv/\spun$ & \mlp  \\ 
\noalign{\smallskip}
\hline\noalign{\smallskip}
RHD              &0.28&1   &$1.723$&$1.577$\\
RHD              &0.28&5   &$1.738$&$1.588$\\
RHD              &0.24&1   &$1.782$&$1.609$\\
RHD ({\it inferred\/})&0.24&5&$\mathit{1.800}$ &$\mathit{1.609}$ \\
\noalign{\smallskip}
%\hline
\noalign{\smallskip}
Helioseismology&0.24&\mbox{-}&$1.778$&$1.727$\\
Helioseismology&0.28&\mbox{-}&$1.736$&$1.594$\\
\noalign{\smallskip}
%\hline
\noalign{\smallskip}
RHD            &0.246&5       &$1.791\pm0.01$& $1.61\mp0.05$\\ 
Helioseismology&0.246&\mbox{-}&$1.771\pm0.005$& $1.71\mp0.02$\\ 
\noalign{\smallskip}
\hline
\end{tabular}%
\end{flushleft}
\label{t:solardata}
\end{table}

Table~1 summarizes the findings for the Sun.  For interpreting the
different entropies given in the table we note here that the entropy
jump over the superadiabtic layers in the Sun amounts to $\approx
0.2\pun{\spun}$.  The first block of data (4 entries) in Table~1 refers
to RHD models.  For solar effective temperature and gravitational
acceleration we calculated RHD models with $\abuhe=0.24$ in addition
to models with our standard $\abuhe=0.28$.  We considered RHD models
where the radiative transfer was treated in grey approximation
($N_\nu=1$), and where its frequency dependence was approximately
taken into account using 5 representative frequency bands ($N_\nu=5$,
for details about the radiative transfer scheme see Ludwig et
al. 1994).  The entropies~$\senv$ of the adiabatic part of the
convective envelopes are based on the assumption $\senv=\sstar$.
The entropies derived from the RHD simulations were converted into an
effective mixing-length parameter~\mlp\ as outlined above.

There is a noticeable dependence of \senv\ on the treatment of the
radiative transfer and the helium content.  The different treatment of
the radiative transfer leads to a different atmospheric temperature
structure, resulting in a change of the absolute entropy level in the
deeper atmosphere.  There is some further influence on the entropy
jump itself since the radiative energy exchange in the surface layers
is slightly modified.  But despite significant differences in \senv\
for the two RHD models with $\abuhe=0.28$, both models give the same
{\mlp} values within the uncertainty of \mlp\ of $\pm 0.05$. This is
achieved by comparing RHD models and envelope models on a strictly
differential basis.  The \ttaurelation\ in the envelope models was
selected to closely follow the relation found in the RHD models:
RHD models basing on grey radiative transfer were compared to
envelope models calculated with a grey \ttaurelation\ in their
atmospheric layers. For RHD models basing on frequency-dependent
radiative transfer, a \ttaurelation\ was derived by horizontally and
temporally averaging their temperature structure on surfaces of
constant optical depth.  This \ttaurelation\ was subsequently used in
the calculation of the related envelope models. When mixing solar RHD
and envelope models with incompatible {\ttaurelation}s, \mlp\ changes
by up to $0.1$. We have performed the same tests for a hotter RHD
($\Teff=6500\pun{K}$) model with similar findings. The insensitivity
of \mlp\ with respect to the treatment of the radiative transfer allowed 
us to largely rely on computationally less demanding RHD models employing
grey radiative transfer for studying the scaling of \mlp\ across the
HRD.
 
The entropy difference due to a change in the helium content is
primarily related to the associated change of the mean molecular
weight.  To first order, the RHD models react to changes in \abuhe\ by
keeping the temperature-pressure structure invariant.  The density
scales in proportion to the change in molecular weight (here:
$\mu(\abuhe=0.28)/\mu(\abuhe=0.24)=1.037$); entropy and internal energy
per unit volume remain the same, allowing the model to transport the same
energy flux without a change of the flow velocity.  The simple scaling
behaviour of the RHD models with \abuhe\ and the invariance of \mlp\
found in the case $\abuhe=0.28$ lead us to derive the values for
$\abuhe=0.24, \mathrm{N}_\nu=5$ from an envelope model {\em
assuming\/} the same \mlp\ as in the case $\abuhe=0.24,
\mathrm{N}_\nu=1$.  The entry for this model in the table is flagged
as ``inferred''.

The second block of data (2 entries) in Table~1 shows results derived
from helioseismic considerations.  Helioseismology provides the sound
speed-density profile in the Sun with high accuracy.  In the deeper,
adiabatically stratified layers of the envelope this profile defines
an adiabat in the sound speed-density plane.  If chemical composition
and the EOS are known one can label this adiabat with its actual
entropy value.  Note that only thermodynamic properties enter into
this procedure; it is independent of opacity or results from stellar
structure models as long as one regards the helioseismic measurements
themselves as independent.  The helioseismic entropy values of the
second block of data in Table~1 are derived in this way by
assuming a certain \abuhe\ and adopting the OPAL~EOS (Rogers
et al. 1996) to establish the relation between sound
speed, density, and entropy.  The {\mlp} values are again derived from
envelope models that match \senv. In contrast to \senv, the determination of
{\mlp} depends on low-temperature opacities and stellar structure 
considerations. The helioseismic data are due to Antia (1996) and were
kindly made available to us by the author.

The third block of data (2 entries) finally summarizes our best
estimates for \senv\ from RHD models and helioseismology.  They were
obtained by linear interpolation between the data for $\abuhe=0.24$
and $\abuhe=0.28$ to Antia's (1996) best helioseismic estimate for the
helium content $\abuhe=0.246$ based on the OPAL~EOS.  The
uncertainties given in the helioseismic case are dominated by the
uncertainty of the helium content.  We have adopted an uncertainty of 
$\pm 0.005$, roughly comprising the range of \abuhe\ discussed in the
literature.  In the RHD case the uncertainties reflect the variations
which we find between simulation runs for the same physical parameters
that differ in numerical details (grid resolution, size of the
computational box, explicit or implicit treatment of the energy
equation, etc.).

Of course, all our error estimates do not include possible systematic
effects. The helioseismic \abuhe\ can be affected by systematic errors
in the EOS. The RHD results are certainly affected by the
two-dimensionality of the models.  Indeed, we interpret the
discrepancy between RHD and helioseismic results as systematic effect:
our 2D RHD models overestimate \senv\ by $0.02\pun{\spun}$ (10\% of
the overall entropy jump between the surface and the adiabatic layers)
and correspondingly underestimate the solar \mlp\ by $0.1$.  This conclusion
is supported by first results from a differential comparison of 2D and 3D
RHD simulations for the Sun.  The 3D run predicts a smaller \senv\ by
$0.013\spun$, corresponding to a larger \mlp\ by $0.07$. Taking this fact
into consideration, the theoretical and observational determination of
\mlp\ become fully consistent.

We expect another systematic effect to influence the absolute value of
\mlp\, affecting both the observationally and the theoretically derived 
values in the same manner.  There exist indications that our low-temperature
opacities underestimate the real stellar opacity in the deeper,
subphotospheric layers.  Our differential approach for determining
\mlp\ cannot fully compensate for this deficiency since the opacities 
affect RHD and MLT models differently. In RHD models the opacity effects are
confined to the very thin cooling layer on top of upflowing regions,
while in MLT models the opacities are important all over the zone of
high superadiabaticity.  This biases our {\mlp} values again towards
lower values.  In the region of high superadiabaticity our opacities
are presumably too small by 10--20\%, causing \mlp\ to be underestimated 
by $\approx 0.1$.

However, in view of the fact that the RHD models are essentially 
parameter-free we consider the match in 2D as satisfactory.  Besides the 
zero point, the functional dependence of \senv\ and \mlp\ on the fundamental
stellar parameters is of major interest. We believe that for these scaling 
relations our error estimates based on purely intrinsic model uncertainties 
are well justified.

\section{Calibration data for solar-type stars}

As for the Sun, we studied the behaviour of \senv\ and \mlp\ over the
HRD in the solar neighbourhood.  This investigation demanded a large
number of simulation runs since we had to cover many objects and had
to clarify how densely the HRD should be sampled in order to record the
relevant variations of \mlp. The results are now based on 58 simulation runs
for solar metallicity and $\abuhe=0.28$, covering the range
$4300\pun{K}\leq\Teff\leq 7100\pun{K}$ and $2.54\leq\logg\leq 4.74$.
Corresponding results for metallicities down to
$\moh=-2$ will be published in a subsequent paper.

Various aspects determined the region of the HRD which is covered by
our models: Towards higher \Teff\ the convective zones become shallow
and hence insignificant for the overall stellar structure. Towards lower
\Teff\ molecular opacities become more and more important; these are
not included in our opacities yet.  Moreover, convection becomes very
efficient, leading to small entropy jumps which cause problems in the
numerical modeling.  Towards lower \logg\ the large entropy jumps
cause other numerical problems related to the steep temperature and
velocity gradients in the surface cooling layer. And finally, 
no solar-type stars are located towards higher \logg.

We present our findings in graphical as well as numerical form.  Like
in the solar case, several RHD simulation runs differing in numerical
details were performed for certain atmospheric parameter combinations,
thus allowing us to estimate the internal uncertainties associated
with our models.  In the figures we provide numerical values and show
polynomial or exponential least-squares fits to the data points.  The
fits are constructed such that the residuals between fitted and actual
data values show no remaining systematic dependence on the stellar
parameters. In Appendix~C we give explicit expressions for these
fits. In this way our results can be easily utilized in other
applications. We made efforts to produce fits with a simple and smooth
functional dependence. Nevertheless we remind the reader to the fact
that polynomial fits of higher order are not well suited for
extrapolation. In other words: the fits quickly loose their meaning outside
the region covered by the grid of RHD models and should not be extrapolated
too far!

\begin{figure*}
\resizebox{\hsize}{!}{\includegraphics[draft=\draftflag]
{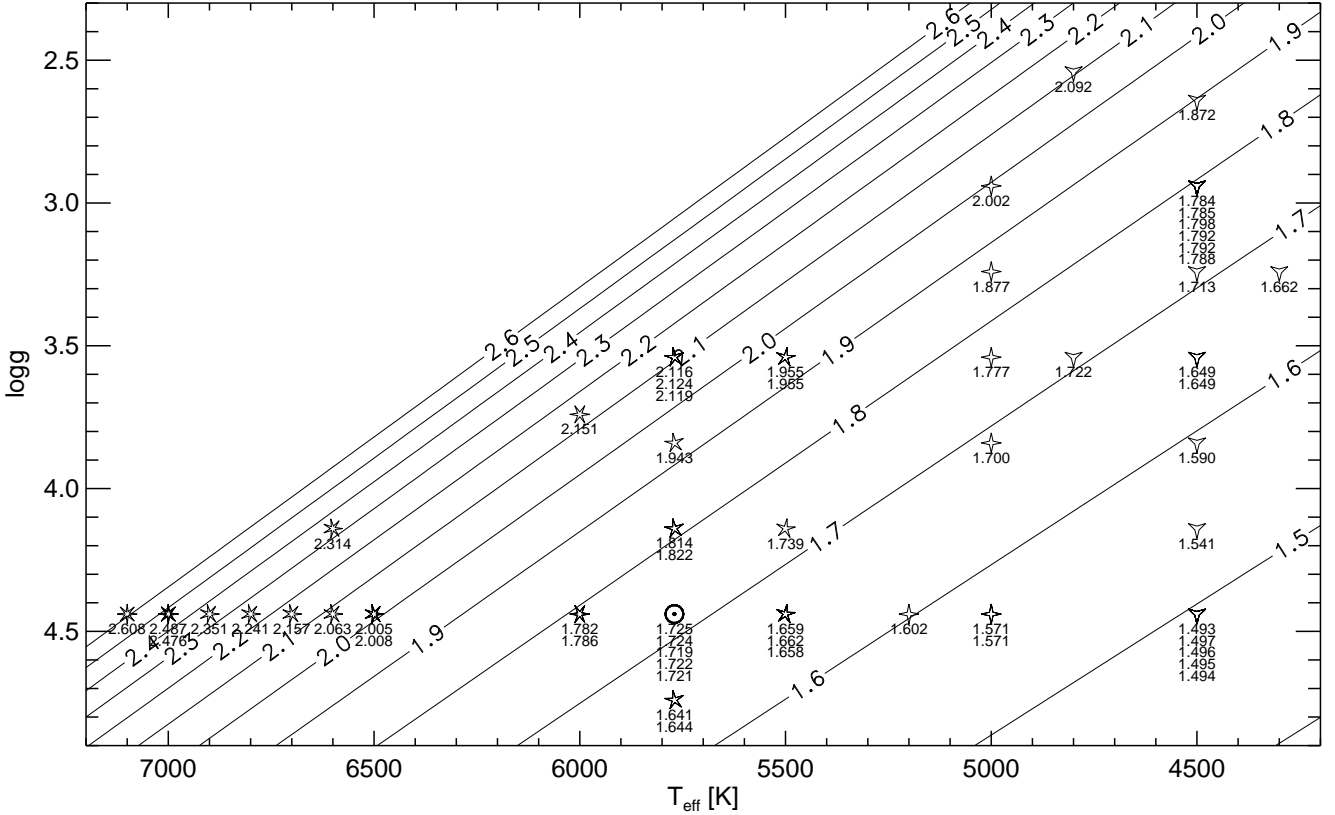}}
\caption[]{%
\senv\ in units of $\spun=10^9\pun{erg/g/K}$ from the grid of RHD models for
$\abuhe=0.28$ and solar metallicity. Symbols indicate RHD models. Attached
to the symbols the actual \senv\ values are given; the contour lines
present a fit to them. The fitting function is given in Appendix~C.
\label{f:sstarhd}
} % end of caption
\end{figure*}

\begin{figure*}
\resizebox{\hsize}{!}{\includegraphics[draft=\draftflag]
{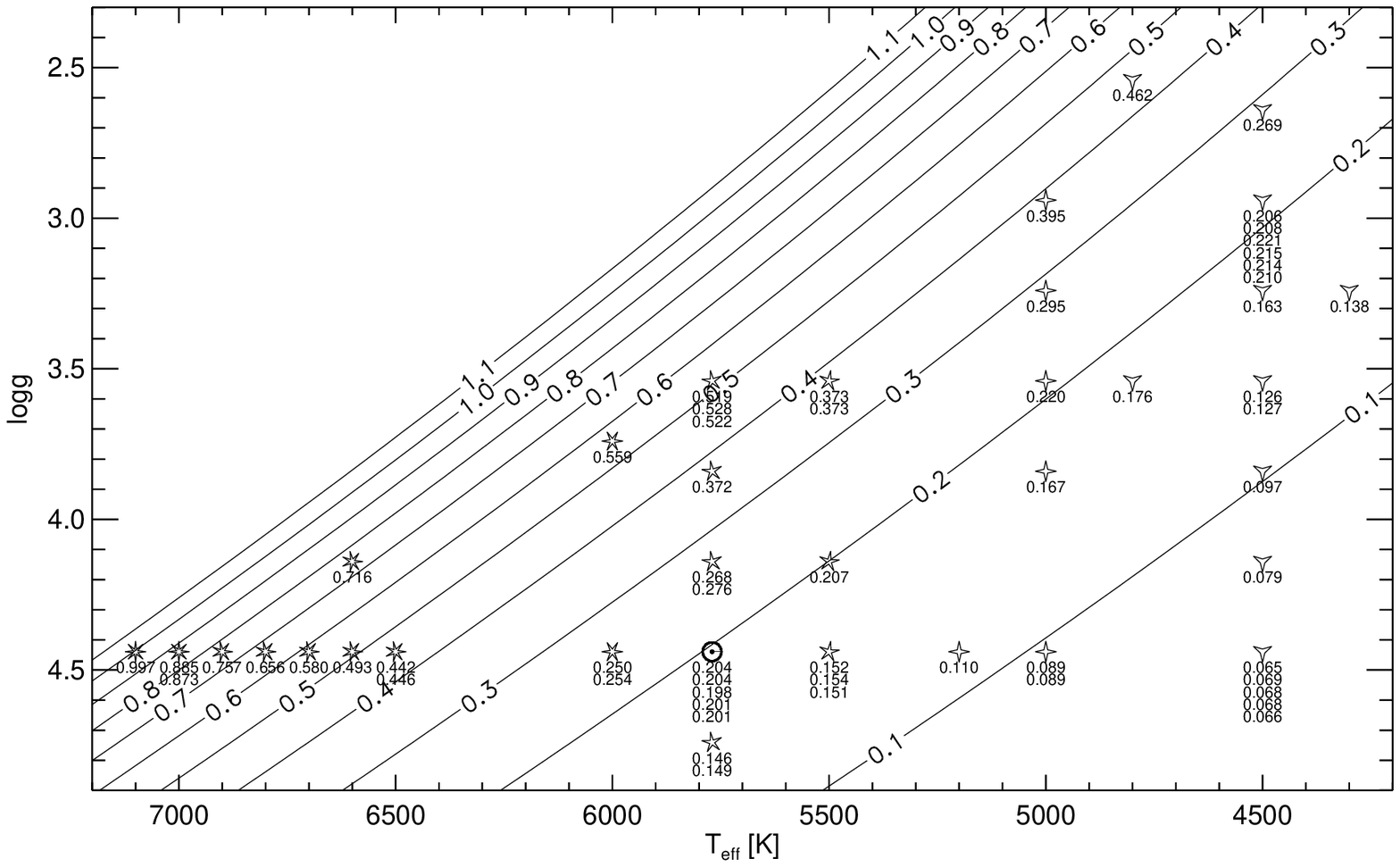}}
\caption[]{%
Entropy {\em jumps\/} in units of $\spun=10^9\pun{erg/g/K}$. The
entropy jump was calculated as difference between \senv\ and the entropy
at the photospheric entropy minimum. See Fig.~\ref{f:sstarhd} for
further explanations.
\label{f:sjumps}
} % end of caption
\end{figure*}

Figure~\ref{f:sstarhd} displays \senv\ as a function of \Teff\ and
\logg\ in the HRD.  In the region where $\senv>2.6\times
10^9\pun{erg/g/K}$, convection becomes inefficient and the convective
zones are superadiabatic throughout.  Entropies in that region refer to
the value encountered at the base of the convective envelope and are
still derived assuming $\senv=\sstar$.  Besides \senv\ itself, the entropy
jumps are of interest since they measure the efficiency of the convective 
energy transport.  The entropy jumps shown in Fig.~\ref{f:sjumps} are the 
entropy differences between the photospheric entropy minimum on the
optical depth scale and \senv\ 
(see Fig.~\ref{f:splots}).

Figure~\ref{f:sjumps} shows that the RHD model grid covers more than an
order of magnitude in the entropy jump, comprising stars with rather
efficient envelope convection (with small entropy jumps) in the
regions of K-dwarfs and stars with inefficient convection (large
entropy jumps) in the region of F-dwarfs.  The significant variation
of the entropy jumps indicates that the simple dependence of \senv\
seen in Fig.~\ref{f:sstarhd} is non-trivial.  It is not just
reflecting the changes of the stellar surface conditions due to
changes of opacity and ionization balance but carries information
about changes in the efficiency of the convective energy transport.

% --- \senv from hydro models
%

% --- \senv from hydro models

\begin{figure*}
\resizebox{\hsize}{!}{\includegraphics[draft=\draftflag]
{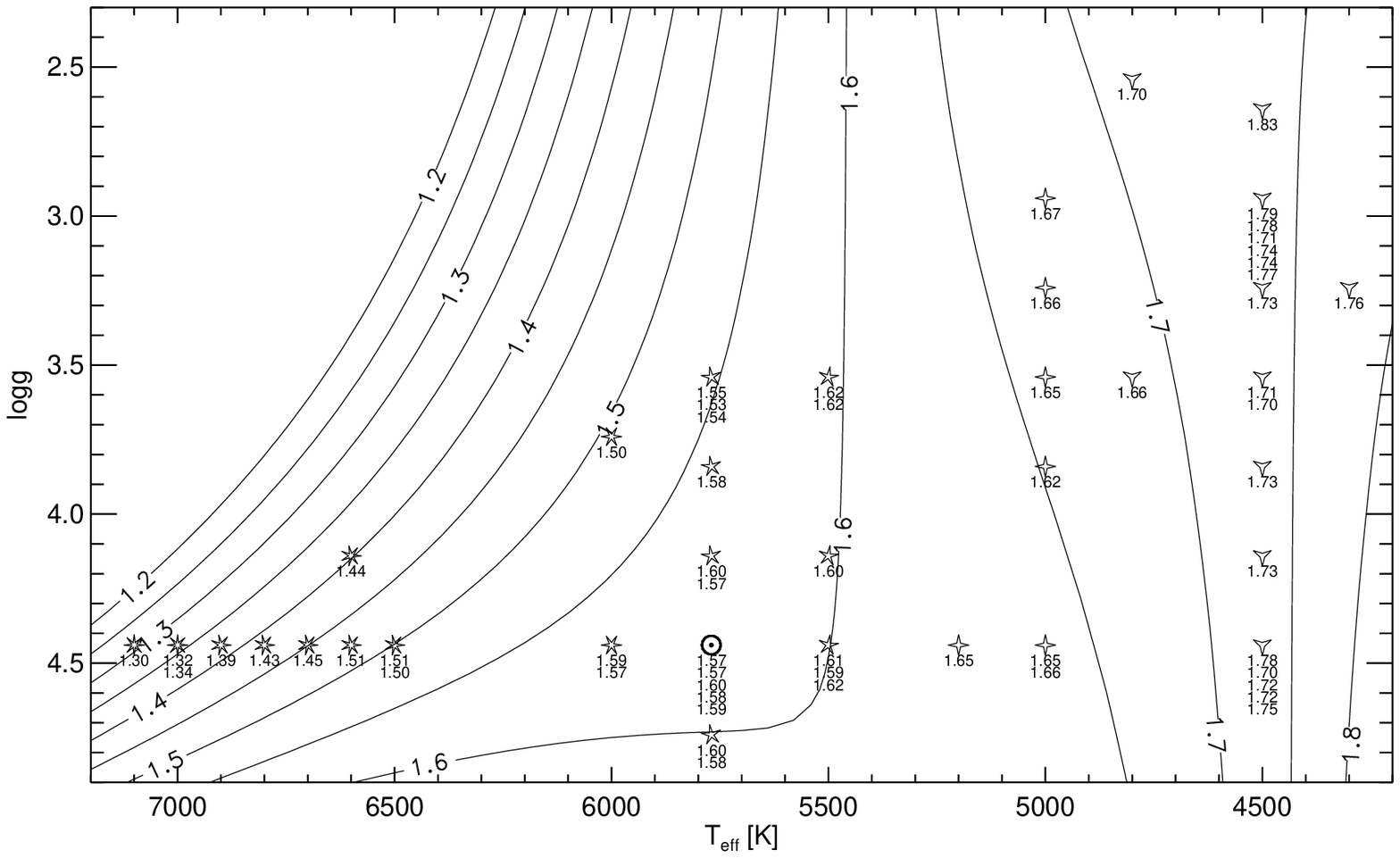}}
\caption[]{%
\mlp\ for standard mixing-length theory (B\"ohm-Vitense 1958) with
$\lmix = \mlp\,\Hp$ (\lmix: mixing-length, \Hp: local pressure scale
height).  The presentation of the data is analogous to
Fig.~\ref{f:sstarhd}.
\label{f:mlp}
} % end of caption
\end{figure*}

With the aid of envelope models (as described before) we translated the
{\senv} values given in Fig.~\ref{f:sstarhd} into equivalent mixing-length
parameters which are shown in Fig.~\ref{f:mlp}.  We find a moderate,
nevertheless significant variation of \mlp\ between about 1.3 for
F-dwarfs and 1.75 for K-subgiants.  In the close neighborhood of the
Sun we find a plateau where \mlp\ remains almost constant.  As
discussed previously, the absolute values of \mlp\ are probably less
reliable than its scaling properties as displayed in Fig.~\ref{f:mlp}.

\begin{figure*}
\resizebox{\hsize}{!}{\includegraphics[draft=\draftflag]
{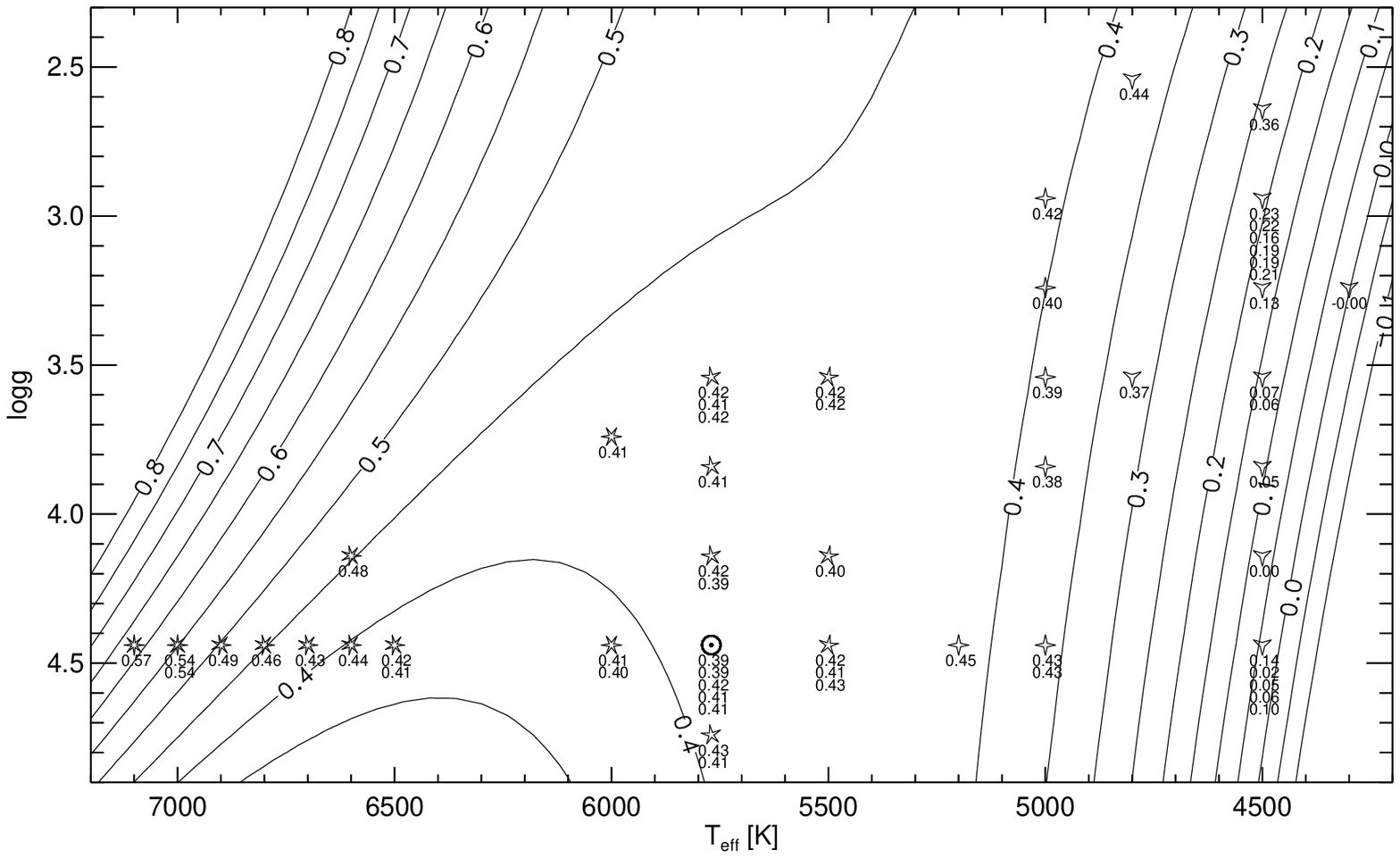}}
\caption[]{%
\mlpcm\ for the Canuto \&\ Mazzitelli convection theory with $\lmix =
z + \mlpcm\,\Hptop$ ($z$: distance to the upper
(Schwarzschild) boundary of the convective zone, \Hptop:
pressure scale height at the upper boundary). The \mlpcm-calibration
reproduces the same underlying \senv-distribution as the
\mlp-calibration shown in Fig.~\ref{f:mlp}. The presentation of the
data is analogous to Fig.~\ref{f:sstarhd}.
\label{f:mlpcm}
} % end of caption
\end{figure*}

The theory of Canuto \&\ Mazzitelli (1991,1992; CMT) describes
convection analytically within the picture of a turbulent medium.  In
view of the significant amount of work that went into the
implementation and verification of this prescription by several
groups, it appeared worthwhile to compare our results with CMT.  Like
MLT, CMT contains a free length-scale $\lmix$.  According to Canuto
\&\ Mazzitelli $\lmix$ is essentially the distance to the upper
(Schwarzschild) boundary of the convective envelope; for purposes of
fine tuning they add a small fraction of the pressure scale height at
the upper boundary.  We follow this recipe here by writing $\lmix = z
+ \mlpcm\,\Hptop$ and translated our \senv\ to \mlpcm\ values, again
with the help of envelope models, but now based on CMT.  As shown in
Fig.~\ref{f:mlpcm}, we find $\mlpcm\approx 0.4$ for the Sun.  To
assess the sensitivity of \senv\ to \mlpcm, we note that the overall
entropy jump at fixed solar \Teff\ and \logg\ increases by roughly a
factor of $2$ if one reduces \mlpcm\ from $0.4$ to zero.  As in the
MLT case, we find a region of rather constant \mlpcm\ in vicinity of
the Sun and a significant variation between zero in the region of
K-dwarfs and almost $0.6$ for F-dwarfs. Interestingly, for the model
with the lowest \Teff, we find an \mlpcm\ slightly smaller than
zero. This value was formally derived by extrapolating the
$\senv(\mlpcm)$-relation towards negative \mlpcm.

% --- \alpha calibration

% --- alpha_CM(z) calibration

\section{Implications and discussion}

%In the following we discuss a --- probably not exhaustive --- list of
%implications.  In this rather subjective selection we emphasize the
%link between our results and stellar evolution modeling.

\subsection{Calibrating \mlp\ with evolutionary models of the Sun?}

The RHD models predict almost no change of \mlp\ during the solar
main-sequence evolution and at most a very weak dependence of the
convective efficiency on the helium abundance in the convective
envelope (see Fig.~\ref{f:mlp} and Table~1). This is an
important piece of information since it justifies the common practice
of calibrating the solar \mlp\ with evolutionary models.  However, we
think that the accurate knowledge about the solar envelope structure from
helioseismology now provides a cleaner way to calibrate the efficiency
of solar convection by the procedure outlined in Sect.~\ref{s:sbm}.
Calibrating \mlp\ with the help of the present envelope structure
would provide insight into the validity of parts of the physics put
into the evolutionary model calculations.

For instance, calibrated solar models with and without element
diffusion, respectively, provide different values for \mlp\ 
(cf. e.g. Richard et al. 1996).  Models including diffusion give a 
lower helium abundance
at the surface for the present Sun and a somewhat higher \mlp\ than
models without diffusion.  This might appear to contradict our
findings, namely that \mlp\ is almost independent of the helium
content.  The reason is of course that the evolutionary models without
diffusion do not represent the Sun since their present surface helium
abundance is not the actual one.  This shortcoming of an evolutionary
model without diffusion is compensated for by a change of \mlp.  As a
price, no assessment of the quality of the model or the adequacy of the
input physics is possible any more. If, however, one calibrated \mlp\ at 
the present envelope structure, shortcomings in the input physics would 
not be hidden in the \mlp\ calibration but would show up as a mismatch 
between the predicted and the actual solar radius.

\subsection{How to make best use of the calibration data in 
            evolutionary models?}

In the case of the Sun our calibration underestimates the absolute
value of the mixing-length parameter by about $0.1\ldots 0.2$, a value
which can be explained by a combined systematic effect due to the too
small low-temperature opacities and the 2D approximation.  The scaling
behaviour of the \mlp\ with effective temperature and surface gravity
is a differential result and will be much less affected by these
systematic shortcomings.  From this perspective it seems that a
scaling of the \mlp\ values presented here by a constant factor
(slightly larger than unity) is still admissible. In the context of
stellar evolutionary models we suggest to calibrate \mlp\ at the
present Sun and use the ratio to \mlp\ from our calibration as a
scaling factor.  
We propose a constant scaling factor since it appears plausible to
compensate for the systematic offset seen in the Sun in a simple,
homogeneous fashion.
For the time being this is the best one can do since the Sun is the
only star where we know \mlp\ with sufficient accuracy to critically
test our results observationally.
However, we emphasize that at the moment the scaling is a well-motivated
modification to our results. We expect to eliminate the necessity of this 
step by improving on the identified systematic shortcomings of our approach 
in future models.

The simplified radiative transfer in the RHD models provides only an
approximate atmospheric \ttaurelation.  We have seen by comparison of
grey and non-grey models that the influence of the \ttaurelation\ on
the determination of \mlp\ can be strongly reduced if one sticks to
the same approximation in the envelope and RHD models.  In this way we
separated to some extend properties of convection from issues related
to opacities and radiative transfer.  If one now aims at a comparison
with observations, the radiative properties should be modeled as
accurately as possible.  In evolutionary models intended to reproduce
observations (e.g. colors) one should use as accurate {\ttaurelation}s as
possible, preferentially derived from full-fledged model atmospheres.
The \mlp\ values from our calibration are not affected by such a
change to a moderately differing \ttaurelation.

\subsection{Consistency with stellar stability}

A variation of \mlp\ and corresponding variation of \senv\ with the
atmospheric parameters \Teff\ and \logg\ can lead to a secular
instability of a star on the Kelvin-Helmholtz timescale of its
convective envelope.  In order to see this, let's consider a small
perturbation of \senv\ in a star's envelope.  As known from stellar
structure models such a disturbance alters its radius while the
luminosity remains largely unchanged. The change in radius leads to a
change of the surface parameters. Following Christensen-Dalsgaard
(1997, his relations (10) and (12)) one approximately finds for the
differential relation between \senv\ and \Teff\ 
valid for stars similar to the Sun
\beq
\left(
\frac{\partial\,\senv}{\partial\,\Teff}\right)_{\mathrm{L=const}} 
\approx -5 \frac{\cp}{\Teff}
\label{e:senvr}
\eeq
where \cp\ is the typical specific heat at constant pressure in the
convective envelope.  

While relation~\eref{e:senvr} describes the response of the global
stellar structure on changes of \senv\ the question is whether this is
compatible with the surface conditions. 
Since from the surface conditions \senv\ is a
function of \Teff\ and \logg, there is the possibility of a feedback on
\senv\ which can amplify or damp an perturbation of \senv.
One would obtain an amplification if \senv\ drops more strongly with
\Teff\ than given by relation~\eref{e:senvr}.  In case of an
amplification, rapid changes of the stellar surface parameters would
occur, leading to regions in the HRD devoid of stars.

Clearly, we do not observe such ``gaps'' in the region of the HRD 
studied here, and indeed one can verify with the aid of
Fig.~\ref{f:sstarhd} that \senv\ as controlled by the surface
conditions exhibits a stabilizing behaviour
since
\beq
\left( \frac{\partial\,\senv}{\partial\,\Teff} \right)_{\mathrm{L=const}} > 0.
\eeq
\noindent
Similar relations hold for the atmospheric entropy minimum (not shown)
and the entropy jump (see Fig.~\ref{f:sjumps}).  Hence, all factors
influencing \senv\ suggest that the convective instability described
above will not be encountered.  However, at least in principle,
it cannot be excluded that under special circumstances a convectively 
driven runaway might occur.

\subsection[]{The {gravity-darkening\/}\footnote{Sometimes also
referred to as gravity-{\em brightening\/}!} exponent}

On the surface of a slowly rotating star the relation between local
\Teff\ and gravity~\grav\ can be approximated by 
\beq
\Teff\propto\grav^\beta.
\eeq
Lucy (1967) argued that the so called gravity-darkening exponent~$\beta$
for stars with convective envelopes is given by 
\beq
\beta = 
\left(\frac{\partial\log\Teff}{\partial\logg}\right)_{\mathrm{\senv=const}}.
\eeq
Figure~\ref{f:sstarhd} represents lines $\senv=\mbox{const}$, so
the gravity-darkening exponent given by our RHD simulations can be
readily deduced from this plot.  We find an increase of $\beta$ from
0.07 to 0.10 when going from the F- to the K-dwarfs, and a slight
decrease of $\beta$ with decreasing gravity.  Basically, this confirms
Lucy's result $\beta=0.08$, with the novelty that our approach
eliminates the weakest point of his analysis, namely that MLT provides
a reasonable scaling relation $\senv(\Teff,\logg)$ with constant \mlp.  
Current observations of $\beta$ (see e.g.  Alencar \&\ Vaz~1997) are
consistent with our findings, but show a rather large scatter and do
not allow a critical test of our results. We therefore refrain from a
further discussion here.

\subsection{Contrasting ours with other approaches}

The long-standing lack of a reliable theory of convection has prompted
numerous attempts to remedy the situation. In the following we comment
on the work of two groups and highlight the major differences in the 
involved physics between their's and our approach, hoping to clarify
the possible reasons for deviating results.

In a series of three papers Lydon et al. (1992, 1993a,b; hereafter
LFS) presented a formulation of convective transport based on results
of numerical experiments by Chan \&\ Sofia (1989).  The idealized
numerical experiments were set up to describe the generic properties
of almost adiabatic convection; radiative energy transport was
included only as a diffusive flux computed with constant
conductivity.  Chan \&\ Sofia extracted {\em local\/} statistical
relations from their numerical data which were subsequently used by
LFS to derive an expression for the convective energy flux suitable 
for calculating stellar
structure models.  LFS modeled the Sun as well as the A- and
B-component of the $\alpha$~Centauri system.  They found essentially
the same \mlp\ in all cases when they translated their formulation
into an effective mixing-length parameter.  This is consistent with
our results, but in our opinion not a strong statement since the rather
small differences in the surface parameters and chemical composition
among the three stars makes it difficult to detect changes of \mlp.
Moreover, there is a principle problem with the LFS convection
formulation since it relies on numerical experiments for adiabatic
convection.  In a follow-up project to the work of Chan \&\ Sofia,
Kim et al. (1995) found significantly different statistical relations
among the fluctuating quantities in the superadiabatic regime, probably
implying that larger uncertainties have to be attributed to the original
results of LFS.

In CMT the main improvement with respect to MLT was the inclusion of a
larger spectrum of eddies supposed to exist in a turbulent medium.
Radiation was only included in a MLT-like fashion, and effects due to
the compressibility of the medium were neglected.  In our
investigation we have regarded the characteristic length
scale in CMT as a free parameter, adjusting it to fit the results of our
simulations.  If we compare the calibration of CMT shown in
Fig.~\ref{f:mlpcm} with standard MLT shown in Fig.~\ref{f:mlp}, we find
a more extended plateau in the solar vicinity.  The general tendency of
CMT to make efficient convection more efficient and inefficient
convection less efficient results in a smoother behaviour relative to MLT.
But towards high and low effective temperatures, a clear
over-compensation occurs, leading to a change of \mlpcm.  In MLT as well
as in CMT, a significant variation of the free parameter is needed, and no
simple scaling is apparent.  We do not find a qualitative difference
in their ability to reproduce \senv\ from the RHD simulations.

LFS as well as CMT essentially relate the convective flux to the {\em
local\/} conditions in the flow.  Such an approach is not well motivated
from the simulation results where convection appears to be an
extremely non-local phenomenon governed by the processes taking place
in the thin cooling layer at the stellar surface (see Spruit 1997 for
a recent discussion) while the bulk of the convection zone adapts to
these layers. Even though we fit our simulation results to local theories,
we do not assume that convection can be described in local terms since
the fit is of purely formal nature.  Furthermore, both LFS and CMT treat the
effects of radiative transfer in a rudimentary fashion, despite the 
well-known fact that for a {\em quantitative\/} description of stellar 
convection a proper treatment of the radiative transfer is crucial. 
Indeed, much effort goes into the realistic modeling of radiative transfer 
in our RHD simulations.  CMT describes
convection within the picture of incompressible turbulence.  However, it
is just the compressibility of the stellar gas that gives 
turbulence in stars its special inhomogeneous character (see Nordlund 
et al. 1997 for a detailed discussion). In contrast, our RHD simulations
fully include the effects of compressibility.

\section{Concluding remarks}

We presented a calibration of the asymptotic entropy~\senv\ and the
corresponding mixing-length parameter~\mlp\ for solar-type stars
basing on radiation-hydrodynamics models. Despite the fact that
\senv\ is all that is needed to construct a stellar structure model,
it was helpful to translate \senv\ into a corresponding \mlp\ since
\mlp\ proved to be less sensitive to the physical and numerical input
to the models than \senv\ itself.  We gave only a description of the
numerical results without providing an explanation in physical terms
--- not even on a qualitative level --- of the scaling behaviour of
\mlp\ which we observe.  We shall come back to this interesting issue
in the next paper of this series where we can look at it from a
broader perspective by including models of sub-solar metallicity.

Figure~\ref{f:sstarhd} shows a remarkably simple structure if one
considers the complex interplay of fluid flow and radiation which
governs the dependence of \senv\ on the stellar parameters.  
Certainly, the simplicity of this dependence is a major reason
for the relative success of MLT to predict \senv, as
evidenced by the moderate variation of \mlp\ in Fig.~\ref{f:mlp}.
Looking at relative changes ignores the more fundamental problem  
of fixing the absolute value
of \mlp, which in practice is done by taking recourse to empirical
constraints.  Our RHD models provide a determination of the zero point
from first principles.  Since in MLT important pieces of the physics
of convection are missing --- at least within the present physical
interpretation of the MLT formulae --- our work should not be
considered as validation of MLT, even though MLT is capable of matching
some of our simulation results quite well.

Work is under way to check and apply the hydrodynamical convection
models beyond
the comparison with the Sun.  There are classical procedures which allow
the determination of the convective efficiency at various locations in
the HRD (position of the red giant branch, shape of the main sequence,
evolution of binary stars).  A lot of work has already been dedicated
to empirical determinations, but conclusions are sometimes
conflicting and no clear picture has emerged yet. We suspect that
systematic uncertainties are actually often larger than
estimated. 
E.g. Castellani et al. (1999) emphasize discrepancies in fitting the
main sequence of open clusters which are related to the
temperature-color transformation and the uncertainties in \mlp.
Clearly, with an independent calibration of \mlp\ at hand one can
disentangle both effects.  For the case of globular clusters Freytag
\&\ Salaris (1999) have studied the effects which are expected from
our calibration on the shape of the turn-off and the position of the
red giant branch.
By using our calibration the effective temperatures of their evolutionary
models become essentially unaffected by the uncertainties inherent to MLT.
The uncertainties related to the temperature-color transformation remain
present but one can at least judge the internal accuracy of the
transformation.
The precise
HIPPARCOS data --- in particular for some open clusters --- might
allow a detailed investigation of these issues.  Moreover,
helioseismology proved to be an invaluable tool in the
case of the Sun, and astero\-seis\-mic measurements of the internal
stellar structure appear to be a promising way to gain further
insight.

Already in 2D, the construction of {\em hydrodynamical model grids\/}
is computationally demanding.  Nevertheless, the step towards 3D models
is desirable, and first calibrations of \mlp\ from 3D models are
now available for smaller stellar samples: Trampedach et
al. (1997) presented a calibration based on 6 models in the solar
vicinity, Abbett et al. (1997) presented a calibration for the Sun.
The latest version of the calibration of Trampedach et al. (Trampedach
1998, priv. comm.) is consistent with our calibration except for a
systematic offset which is caused by differences in the employed 
low-temperature opacities; the scaling behaviour is similar as far as it
is possible to judge from the small set of models.  Abbett et al. do
not provide \mlp\ with sufficient precision to allow a critical comparison
with our results.  Own work is in progress to quantify systematic
differences between 2D and 3D models.  As already mentioned, first
results for the Sun give a change of \mlp\ by $+0.07$ with respect to 
the 2D models. We expect systematic changes of similar magnitude
for other stars.  Hence, it appears unlikely that conclusions of this
paper will have to be altered substantially when more 3D results become 
available.

Last but not least we want to reiterate the warning that the presented
calibration of \mlp\ is only intended to reproduce \senv\ and the
entropy jump.
The detailed temperature profile of the superadiabatic layers is not 
necessarily represented adequately by
an MLT model with our calibrated \mlp\ (see Fig.~\ref{f:splots}).  
Moreover, the calibration
is not suitable for providing the optimum mixing-length parameter for
convective stellar atmospheres.  Preliminary results for the Sun
(Steffen \&\ Ludwig~1999) show
that matching the {\em emergent radiation field\/} of multidimensional
simulations with 1D standard atmospheric models requires a quite
different mixing-length parameter than is needed for matching the entropy 
jump. We do not consider this as a contradiction; it just indicates
shortcomings of MLT and the 1D idealization.

\begin{acknowledgements}
During the course of the project, many colleagues helped with
discussions, opinions, and criticism. HGL is particularly grateful to
Norbert Langer, Hendrik Spruit, Maurizio Salaris, Scilla
Degl'Innocenti, Achim Weiss, Philipp Podsiadlowski, Hans Ritter, and
Regner Trampedach. BF acknowledges financial support by the Deutsche
Forschungsgemeinschaft under grant Holweger 596/36-1. MS expresses his
thanks to the University of Kiel for its kind hospitality and for providing
access to its computing facilities.
\end{acknowledgements}

\appendix

\section{Mixing-length formulations}\label{s:mlt}

Generically, the well-known equations of MLT (cf. Cox \&\ Giuli 1968) 
for the convective efficiency
\beq
\Gamma \equiv \frac{\Vact-\Veddy}{\Veddy-\Vad} 
  = \frac{\rho\cp\vconv\taueddy}{f_3\sigma T^3} 
    \left( 1 + \frac{f_4}{\taueddy^2}\right),
\eeq
the convective velocity
\beq
\vconv^2 = f_1 \frac{\lmix^2 g \delta (\Vact-\Veddy)}{\Hp},
\eeq
and the convective flux
\beq
\Fconv = f_2 \frac{\rho\cp\vconv T \lmix (\Vact-\Veddy)}{\Hp}
\eeq
contain four dimensionless free parameters $f_1$,
$f_2$, $f_3$, and $f_4$. In the equations, the optical thickness 
of a convective eddy~\taueddy\ and the isobaric expansion 
coefficient~$\delta$ are defined as
\beq
\taueddy\equiv\chi\rho\lmix, \hspace{1em} 
\delta\equiv -\left(\pdx{\ln\rho}{\ln T}\right)_\mathrm{P}.
\eeq
Further quantities entering the equations are 
the mixing-length~\lmix, specific heat at
constant pressure~\cp, Stefan-Boltzmann's constant~$\sigma$,
density~$\rho$, temperature~$T$, actual temperature gradient~\Vact,
temperature gradient of a convective eddy~\Veddy, and adiabatic
gradient~\Vad\ ($\nabla$ stands for $\partial\ln T/\partial\ln P$).  
Table~\mbox{A1} provides the f-parameters for
various MLT formulations.  The values of \mlp\ given in this
paper refer to the MLT formulation by B\"ohm-Vitense (1958) as
specified in the first row of the table.

\begin{table}
\begin{flushleft}
\caption[]{%
Constants $f_\mathrm{i}$ for the MLT formulations of various authors. (ML1
and ML2 are commonly used terms in the white dwarf community.)}
\begin{tabular}{lllll}
\hline\noalign{\smallskip}
Formulation      & $f_1$  & $f_2$ & $f_3$ & $f_4$  \\ 
\noalign{\smallskip}
\hline\noalign{\smallskip}
B\"ohm-Vitense 1958, ML1 
  & $\frac{1}{8}$ & $\frac{1}{2}$ & 24 & 0 \\
ML2
  & 1           & 2           & 16 & 0 \\
Mihalas 1978, Kurucz 1979
  & $\frac{1}{8}$  & $\frac{1}{2}$ & 16 & 2 \\
Henyey et al. 1965
  & $\frac{1}{8}$  & $\frac{1}{2}$ & $\frac{4\pi^2}{24}$ & $\frac{4\pi^2}{3}$\\
\noalign{\smallskip}
\hline
\end{tabular}%
\end{flushleft}
\label{t:mltfs}
\end{table}

A fine point in our implementation of MLT concerns the evaluation of
the radiative gradient~\Vrad. We do not use the commonly adopted
expression for the radiative gradient in diffusion approximation, but
instead derive \Vrad\ by differentiation of the \ttaurelation. As a
starting point, consider the \ttaurelation\ for a grey atmosphere in 
radiative equilibrium
\beq
T^4 = \frac{3}{4} \Teff^4 \left[\tau + q(\tau)\right].
\label{A5}
\eeq
While in the case of a grey atmosphere $q$ is the Hopf-function, we can
introduce a modified $q$ such that Eq.\,\ref{A5} represents any prescribed 
\ttaurelation.  A reasonable $q$ should have the property
\beq
\lim_{\tau\rightarrow\infty} q = \mathrm{const}
\eeq
which guarantees that \Vrad\ becomes independent of $\tau$ and
consistent with the diffusion approximation for large optical depth
(provided the Rosseland scale is adopted).  The advantage of this
procedure is that we obtain a smooth and realistic \Vrad\ which is consistent
with the specified \ttaurelation\ irrespective of optical
depth.  We never have to distinguish between atmosphere and optically
thick layers and can use MLT throughout, for the small price of having
to compute the $\tau$-scale as well.

\section{Entropy scale}\label{s:sscale}

\begin{table}
\begin{flushleft}
\caption[]{Thermodynamic quantities as a function of entropy and
pressure from the RHD~EOS for solar metallicity and $\abuhe=0.28$. Unless
noted otherwise the values are given in cgs-units.}
\begin{tabular}{rrrrrrr}
\hline\noalign{\smallskip}
$s/\spun$ & $\log P$ & $\log T$ & $\log\rho$ & $\Gamma_1$ & $\delta$ & $\cp/10^8$ \\ 
\noalign{\smallskip}
\hline\noalign{\smallskip}
\noalign{\smallskip}
1.50 &  8.0 &    4.310 &   -4.218 &    1.277 &    1.980 &   10.146  \\
1.50 &  9.0 &    4.487 &   -3.457 &    1.355 &    1.915 &    9.000  \\
1.50 & 10.0 &    4.715 &   -2.746 &    1.467 &    1.616 &    6.608  \\
1.50 & 11.0 &    5.022 &   -2.096 &    1.591 &    1.276 &    4.435  \\
1.50 & 12.0 &    5.375 &   -1.473 &    1.631 &    1.182 &    4.006  \\
1.50 & 13.0 &    5.759 &   -0.868 &    1.663 &    1.104 &    3.602  \\
1.75 &  8.0 &    4.406 &   -4.415 &    1.316 &    1.948 &   11.121  \\
1.75 &  9.0 &    4.618 &   -3.684 &    1.436 &    1.468 &    6.733  \\
1.75 & 10.0 &    4.933 &   -3.032 &    1.577 &    1.124 &    4.074  \\
1.75 & 11.0 &    5.283 &   -2.401 &    1.634 &    1.072 &    3.727  \\
1.75 & 12.0 &    5.675 &   -1.797 &    1.665 &    1.023 &    3.406  \\
1.75 & 13.0 &    6.074 &   -1.197 &    1.667 &    1.015 &    3.373  \\
2.00 &  8.0 &    4.531 &   -4.616 &    1.415 &    1.320 &    6.370  \\
2.00 &  9.0 &    4.859 &   -3.966 &    1.534 &    1.081 &    4.169  \\
2.00 & 10.0 &    5.201 &   -3.325 &    1.646 &    1.026 &    3.520  \\
2.00 & 11.0 &    5.598 &   -2.723 &    1.666 &    1.004 &    3.352  \\
2.00 & 12.0 &    5.998 &   -2.123 &    1.667 &    1.003 &    3.344  \\
2.00 & 13.0 &    6.398 &   -1.523 &    1.667 &    1.002 &    3.342  \\
2.25 &  8.0 &    4.789 &   -4.899 &    1.474 &    1.090 &    4.641  \\
2.25 &  9.0 &    5.124 &   -4.249 &    1.655 &    1.010 &    3.419  \\
2.25 & 10.0 &    5.523 &   -3.648 &    1.667 &    1.001 &    3.341  \\
2.25 & 11.0 &    5.923 &   -3.048 &    1.667 &    1.000 &    3.339  \\
2.25 & 12.0 &    6.323 &   -2.448 &    1.667 &    1.000 &    3.338  \\
2.25 & 13.0 &    6.723 &   -1.848 &    1.667 &    1.000 &    3.338  \\
2.50 &  8.0 &    5.049 &   -5.174 &    1.660 &    1.004 &    3.377  \\
2.50 &  9.0 &    5.448 &   -4.574 &    1.667 &    1.000 &    3.338  \\
2.50 & 10.0 &    5.848 &   -3.974 &    1.667 &    1.000 &    3.338  \\
2.50 & 11.0 &    6.248 &   -3.374 &    1.667 &    1.000 &    3.338  \\
2.50 & 12.0 &    6.648 &   -2.774 &    1.667 &    1.000 &    3.338  \\
2.50 & 13.0 &    7.048 &   -2.174 &    1.667 &    1.000 &    3.338  \\
\noalign{\smallskip}
\hline
\end{tabular}%
\end{flushleft}
\label{t:sscale}
\end{table}

The entropy is not always available from an EOS, and if so it involves
an arbitrary additive constant.  In order to relate
entropy values given in this paper to more common thermodynamic
variables, we provide a small list of quantities as a function of
entropy and pressure in Table~B1, calculated from the EOS
which was used in the RHD simulations.  We have concentrated on higher
temperatures characteristic of the deeper layers of solar-type
stellar envelopes.  The data should allow to interrelate values for
\senv\ to pressures, temperatures, and densities, and provide the
entropy zero point of our EOS.  For interpolation purposes one might
take advantage of the following differential thermodynamic relations
\beq
d\ln\rho =  \frac{1}{\Gamma_1}\,d\ln P - \delta\,\frac{d s}{\cp},
\eeq
and
\beq
d\ln T = \Vad\,d\ln P + \frac{d s}{\cp}.
\eeq
\Vad\ is related to the thermodynamic derivatives given
in Table~B1 according to
\beq
\Vad = \frac{P\delta}{\rho\cp T}.
\eeq

\section{Fitting functions}\label{s:fits}

\begin{table}
\begin{flushleft}
\caption[]{Coefficients for fitting functions}
\begin{tabular}{lrrrr}
\hline\noalign{\smallskip}
coefficient   & $\senv/\spun$ & $\Delta s/\spun$ & \mlp & \mlpcm \\ 
\noalign{\smallskip}
\hline\noalign{\smallskip}
\noalign{\smallskip}
$a_0$    &  1.6488  &  0.029  &  1.587   &  0.414 \\
$a_1$    &  0.0740  &  0.166  & -0.054   & -0.100 \\
$a_2$    & \mbox{}  & \mbox{} &  0.045   & -0.026 \\
$a_3$    &  1.7860  &  1.225  & -0.039   & -0.065 \\
$a_4$    & -1.6762  & -1.209  &  0.176   & -0.092 \\
$a_5$    &  0.1274  &  0.073  & -0.067   &  0.193 \\
$a_6$    & -0.1412  & -0.132  &  0.107   & -0.125 \\
\noalign{\smallskip}
\hline
\end{tabular}%
\end{flushleft}
\label{t:cfit}
\end{table}

The fits for \senv, entropy jump~$\Delta s$, \mlp, and \mlpcm\ shown
in Figs.~\ref{f:sstarhd}, \ref{f:sjumps}, \ref{f:mlp}, and
\ref{f:mlpcm} were computed from expressions~\eref{e:fit3},
\eref{e:fit4}, and \eref{e:fit5} (for \mlp\ and \mlpcm) in terms of
the auxiliary variables \Teffref\ and \loggref\ defined below.  The
coefficients~$a_i$ are listed in Table~C1.  The fitting functions were
generated automatically from a prescribed set of basis functions with
the mere intention of providing numerical fits to the RHD data points
which then allow the RHD results to be utilized easily in other
contexts.  Physical considerations did not enter, and we present the
formulae in a form more suitable for computing rather than for human
interpretation.
We warn the reader that the fits quickly loose their meaning outside
the region covered by the RHD model grid and should not be
extrapolated too far.

\beq
\Teffref\equiv(\Teff-5770)/1000
\label{e:fit1}
\eeq
\beq
\loggref\equiv\log(g/27500)
\label{e:fit2}
\eeq

\beq
\frac{\senv}{\spun} =  a_0 +  a_5 \Teffref +  a_6 \loggref
  +  a_1 \exp[a_3 \Teffref + a_4 \loggref]
\label{e:fit3}
\eeq

\beq
\frac{\Delta s}{\spun} = a_0 +  a_1 \exp[(a_3 +  a_5 
            \Teffref +  a_6 \loggref) \Teffref +  a_4 \loggref]
\label{e:fit4}
\eeq

\beq
\alpha = a_0 +  (a_1 +  (a_3 +  a_5 \Teffref +  a_6 \loggref) 
       \Teffref +  a_4 \loggref) \Teffref +  a_2 \loggref
% \mlp & = &  a_0 +  (a_1 +  (a_3 +  a_5 \Teffref +  
%            (a_6 +  a_7 \Teffref \nonumber\\
%      & + & (a_8 +  a_9 \Teffref +  a_{10} \loggref) 
%            \loggref) \loggref) \Teffref +  a_4 \loggref)  
%   \Teffref +  a_2 \loggref
\label{e:fit5}
\eeq

\end{document}